\documentstyle[epsf,sprocl,pictex]{article}

\font\eightrm=cmr8

\def\Journal#1#2#3#4{{#1} {\bf #2}, #3 (#4)}

\def\NPB{{\em Nucl. Phys.} B}
\def\PLB{{\em Phys. Lett.}  B}
\def\PRL{\em Phys. Rev. Lett.}
\def\PRD{{\em Phys. Rev.} D}

\def\be{\begin{equation}}
\def\ee{\end{equation}}
\def\bea{\begin{eqnarray}}
\def\eea{\end{eqnarray}}

\begin{document}

{\hfill SNUTP 98-141,\ KIAS-P98047}\\

\title{Axion Theory Review
\footnote{Talk presented at Dark Matter 98, Buxton, England,
Sep. 7--11, 1998.}
}

\author{JIHN E. KIM}

\address{School of Physics, Korea Institute for Advanced Study,
207-43 Cheongryangri-dong,\\ Seoul 130-012, Korea, and\\
Center for Theoretical Physics, Seoul National University,
Seoul 151-742, Korea\\ 
E-mail: jekim@phyp.snu.ac.kr} 

\maketitle\abstracts{ I review the axion physics 
with emphases on the couplings
of the very light axion and a possible realization 
in superstring models.}

\section{Why Axions?}

Before 1975, QCD was described by the Lagrangian
$$
{\cal L}=-\frac{1}{2g^2}{\rm Tr}F_{\mu\nu}F^{\mu\nu}
+\bar q(i\gamma^\mu D_\mu-M_q)q.
$$
But after 1975, the following $\theta$ term is known to be present
due to the discovery of instanton solutions in nonabelian
gauge theories \cite{bpst,cdg}
\begin{equation}
\frac{\bar\theta}{16\pi^2}~{\rm Tr}F_{\mu\nu}\tilde F^{\mu\nu}
\end{equation}
where $\bar\theta=\theta_{\rm QCD}+\theta_{\rm QFD}$ which violates
CP. From the experimental bound on the neutron electric dipole
moment \cite{NEDM}, $|\bar\theta|$ is required to be less than $10^{-9}$.
This is $\lq\lq$the strong CP problem".

The axion solution of the strong CP problem is to introduce 
a dynamical field axion ($a$) so that $\bar\theta$ is settled
very near 0.\cite{rev} In this case, one can study the following
Lagrangian below the electroweak scale, after integrating out
the quark fields,
\begin{equation}
{\cal L}=-\frac{1}{4g^2}F^a_{\mu\nu}F^{a\mu\nu}+\bar q
(i\gamma^\mu D_\mu-M_q)q+\frac{a}{32\pi^2F_a}F^a_{\mu\nu}\tilde
F^{a\mu\nu}+\frac{1}{2}\partial^\mu a\partial_\mu a.
\end{equation} 
Here, $\bar\theta$ turns out to be $\bar\theta=a/F_a$ where
$F_a$ is called the axion decay constant. In this model, $\bar\theta$
is a dynamical field. But, for a moment consider $\bar\theta$ as
a parameter of the theory. Denoting
\begin{equation}
\{F\tilde F\}\equiv \frac{1}{32\pi^2}F^a_{\mu\nu}\tilde F^{a\mu\nu},
\end{equation}
we obtain the generating functional in Euclidian space, with
CP violation introduced only in the $\bar\theta$ term, as
\begin{equation}
Z\propto \int[dA_\mu]\prod {\rm Det}(\gamma^\mu D_\mu+m_i)
\exp\left(-\int d^4x[\frac{1}{4g^2}F^2-i\bar\theta\{F\tilde F\}]\right).
\end{equation}
Noting that Det~$(\gamma^\mu D_\mu+m_i)>0$, one can prove
the following inequality,\cite{vw}
\begin{eqnarray}
&e^{-\int d^4x V[\bar\theta]}\equiv
\left|\int [dA_\mu]\prod {\rm Det}(\gamma^\mu D_\mu+m_i)
e^{-\int d^4x(\frac{1}{4g^2}F^2-i\bar\theta\{F\tilde F\})}
\right| \nonumber\\
&\le\int[dA_\mu]\left|\prod {\rm Det}(\gamma^\mu D_\mu+m_i)
e^{-\int d^4x(\frac{1}{4g^2}F^2-i\bar\theta F\tilde F)}\right|\nonumber\\
&=\left|\int [dA_\mu]\prod {\rm Det}(\gamma^\mu D_\mu+m_i)
e^{-\int d^4x\frac{1}{4g^2}F^2}\right|\\
&=e^{-\int d^4xV[0]}\nonumber
\end{eqnarray}
where we used Schwarz's inequality. Thus, $V[\bar\theta]\ge V[0]$
for any $\bar\theta$ without CP violation introduced in the
other sector. As a coupling, any $\bar\theta$ will be good, as
any value of $\alpha_{em}$ is theoretically allowable.

The axion solution is to make $\bar\theta$ a dynamical field
$\bar\theta=a/F_a$. Namely, $a$ has a kinetic energy term, without
any term in the potential except the one coming from the
gluon anomaly term in Eq.~(2). It is possible to introduce 
this kind of pseudoscalar as a Goldstone boson. 
A Goldstone boson nicely fits into this
scheme, since it does not have a potential. But for this
interpretation to work, the current corresponding to the Goldstone
boson must have an anomaly so that $(a/F_a)F\tilde F$ arises, which
is called the Peccei-Quinn mechanism.\cite{pq}  

In addition to the decay constant $F_a$, there is another
fundamental number for axion: the domain wall number 
$N_{\rm DW}$.\cite{domain} It arises because 
$\bar\theta$ is periodic with the period 
$2\pi$ but the fields in the theory can have a phase 
$\bar\theta/N_{\rm DW}$, which can be represented as
$a$ and has a periodicity
\begin{equation}
a\equiv a+2\pi N_{\rm DW}F_a.
\end{equation}
In the standard Big Bang cosmology, there will be a domain wall
problem. But with the inflation with the reheating temperature below
$F_a$, the domain wall problem disappears.  

Remembering that the axion is a dynamical $\bar\theta$, we can easily
derive its interaction terms. For this, we rely on the effective
field theory or calculate them explicitly in a given model.  

As an introduction of axion, let us consider the simplest axion
example, the heavy quark axion or KSVZ axion.\cite{ksvz}
The heavy quark obtains mass by the VEV of a singlet complex 
Higgs field $\sigma$, and the Yukawa coupling is
\begin{equation}
{\cal L}_{\rm Y}=\sigma\bar Q_RQ_L+{\rm h.c.}.
\end{equation}
The potential contains terms invariant under the following
Peccei-Quinn transformation,
\begin{equation}
Q_L\rightarrow e^{-i\alpha/2}Q_L, Q_R\rightarrow e^{i\alpha/2}Q_R,
\sigma\rightarrow e^{i\alpha}\sigma,\ \ \theta\rightarrow\theta-\alpha.
\end{equation}
This symmetry is broken by the VEV of $\sigma$, $<\sigma>=F_a/\sqrt{2}$.
This produces a Goldstone boson, which is
hidden in $D_\mu \sigma^* D^\mu\sigma$
in the original Lagrangian, whose kinetic energy term is
\begin{equation}
\frac{1}{2}\left(1+{\rho\over F_a}\right)^2\partial_\mu a\partial^\mu a
\end{equation}
where $\sigma={(F_a+\rho)e^{ia/F_a}}/\sqrt{2}$. Due to the nontrivial
Peccei-Quinn transformation of the heavy quark, the corresponding
current is not conserved at one-loop level but
\begin{equation}
\partial_\mu J^\mu=\frac{1}{16\pi^2}F_{\mu\nu}^a\tilde F^{a\mu\nu}
\end{equation}
where $F^a_{\mu\nu}$ is the gluon field strength. Thus, the Goldstone
boson obtains mass through this anomalous term.

There are basically two methods to introduce the axion at a scale
$\mu$:\\
\indent (i) As a global symmetry, and\\
\indent (ii) as a fundamental field.\\
But at low energy these two methods give the same result with the
axion coupling as given in Eq.~(3). Case (i) implies the Peccei-Quinn
symmetry.\cite{pq} When this symmetry is spontaneously broken, 
there appears the axion. Above the symmetry breaking scale
$F_a$, there is just the global symmetry. Below the symmetry breaking
scale, there appears the axion--the pseudo-Goldstone boson.
Its fundamental nature can be either
the pseudoscalar present in the Higgs field,\cite{ww,ksvz,dfsz} 
or a composite axion.\cite{comp} Case (ii) involves
a pseudoscalar in gravity theory, including the superstring
axion.\cite{witten} In this case, the decay constant is the
gravity scale $M_{\rm compactification}$. 

The axion properties depend on $F_a$, $N_{DW}$, and couplings to
matter. These couplings drastically differ for diffent axion
models.

\section{Axion Mass and Couplings}

Below the scale $F_a$, we consider light fields of quarks,
gluons, $a$.  This is enough for the study of axion mass.
The axion and gluon Lagrangian is
\begin{equation}
{\cal L}=\frac{1}{2}(\partial_\mu a)^2+({\rm derivative
\ terms\ of\ a})+\left(\theta+\frac{a}{F_a}\right)\{F\tilde F\}
\end{equation}
from which we can redefine the axion field so that the coefficient
of $\{F\tilde F\}$ becomes $a/F_a$. In the KSVZ model, we minimally
created this term at the high energy scale. The low energy physics
at 1 GeV will include quarks. After introducing 
a warm-up example of one-flavor
QCD, we will present the axion mass in the two-flavor QCD.

\subsection{One flavor}

The strong interaction Lagrangian is
\begin{equation}
{\cal L}=-(m_u\bar u_Ru_L+{\rm h.c.})+\frac{a}{F_a}\{F\tilde F\}
+(\rm K.E.\ terms)
\end{equation}
which has the following fictitious chiral symmetry,
\begin{equation}
u_L\rightarrow e^{i\alpha}u_L,\ u_R\rightarrow e^{-i\alpha}u_R,\ 
m_u\rightarrow e^{-2i\alpha}m_u,\ a\rightarrow a+2\alpha F_a.
\end{equation}
There is no chiral symmetry due to the mass term, but we considered
this fictitious symmetry to track down the parameter (here $m_u$)
dependence of the effective potential below the QCD chiral symmetry
breaking scale. Below the chiral symmetry breaking scale, $<\bar uu>
= v^3$, we can write an effective potential consistent with the
symmetry~(13),
\begin{eqnarray}
&V=\frac{1}{2}m_u\Lambda^3e^{i\theta}-\frac{1}{2}\lambda_1\Lambda v^3e^{i(
\eta/v-\theta)}-\frac{1}{2}\lambda_2m_uv^3e^{i\eta/v}\nonumber\\
&+\lambda_3m_u^2\Lambda^2e^{2i\theta}+\lambda_4\frac{v^6}{\Lambda^2}
e^{2i(\eta/v-\theta)}+\cdots+{\rm h.c.}
\end{eqnarray}
where $\Lambda$ is the QCD scale, $\theta=a/F_a$, 
and $\lambda_i$ are couplings.  Assuming that the minimum is 
at $<a>=<\eta>=0$, we obtain the following mass matrix,
neglecting O($m^2_u$) and O($e^{2i\theta}$) terms,
\begin{equation}
M^2=\left(\matrix{\lambda\Lambda v+\lambda^\prime m_uv,\ \ -
\frac{\lambda\Lambda v^2}{F_a}\cr
-\frac{\lambda\Lambda v^2}{F_a},\ \ -\frac{m_u\Lambda^3}{F_a^2}
+\frac{\lambda\Lambda v^3}{F_a^2}}\right)
\end{equation}
where $\lambda$ and $\lambda^\prime$ are newly defined couplings.
Then it has the following determinant
\begin{equation}
{\rm Det}M^2=\frac{m_u\Lambda v}{F_a^2}(\lambda\lambda^\prime v^3
-\lambda\Lambda^3-\lambda^\prime m_u\Lambda^2).
\end{equation}
For $F_a\gg $ (other mass parameters), we obtain $M^2_\eta
=(\lambda\Lambda+\lambda^\prime m)v$ and the axion mass
\begin{equation}
m^2_a=\frac{m_u\Lambda}{F^2_a}\left(\frac{\lambda\lambda^\prime
v^3}{\lambda\Lambda +\lambda^\prime m_u}-\Lambda^2\right).
\end{equation}
If this turns out to be negative, we are in the wrong vacuum, and
should choose $\theta=\pi$. The above axion mass formula shows the
essential feature: it is suppressed by $F_a$, multiplied by the
quark mass, and the rest is the strong interaction parameter.

\subsection{Two flavor}

For the fwo flavor case, we have the following $U(1)_u
\times U(1)_d$ chiral symmetry
\begin{eqnarray}
&u_L\rightarrow e^{i\alpha}u_L,\ \ u_R\rightarrow e^{-i\alpha}u_R,\ \ 
m_u\rightarrow e^{-2i\alpha}m_u,\nonumber\\
&d_L\rightarrow e^{i\beta}d_L,\ \ d_R\rightarrow e^{-i\beta}d_R,\ \ 
m_d\rightarrow e^{-2i\beta}m_d,\\
&\theta\rightarrow\theta+2(\alpha+\beta).\nonumber
\end{eqnarray}
Following the same procedure as before, we diagonalize $3\times
3$ mass matrix, and obtain
\begin{equation}
m_a=\frac{m_{\pi^0}F_\pi}{F_a}\frac{\sqrt{Z}}{1+Z}
\simeq 0.6\times 10^7\left({{\rm GeV}\over F_a}\right)\ {\rm eV}
\end{equation}
where $Z=m_u/m_d\simeq 0.56$.

\subsection{The KSVZ model}

The Yukawa coupling has the form $\sigma\bar Q_RQ_L$ where $\sigma$
is a singlet Higgs field and $Q$ is a heavy quark. Below the scale
$F_a$, the interaction Lagrangian is given in Eq.~(9). The axion current 
is
\begin{equation}
J_\mu=v\partial_\mu a-\frac{1}{2}\bar Q\gamma_\mu\gamma_5Q
+\frac{1}{2(1+Z)}(\bar u\gamma_\mu\gamma_5u+Z\bar d\gamma_\mu
\gamma_5 d)
\end{equation}
where the last term is added in the process of $a,\pi^0,\eta$
diagonalization.

\subsection{The DFSZ model}

The Lagrangian is supposed to be
\begin{equation}
{\cal L}=\sigma\sigma H_1H_2+\bar u_Ru_LH_2+\bar d_Rd_LH_1+
\bar e_Re_LH_1+\cdots+ {\rm h.c.}
\end{equation}
where most of $a$ resides in the phase of $\sigma$, but
$H_1$ and $H_2$ contain small components of $a$. Thus there exists 
the tree level axion-electron-electron coupling, $(2x/(x+x^{-1}))
(a/F_a)m_e\bar ei\gamma_5e$, where $x=<H_2^0>/<H_1^0>$. The current is
\begin{equation}
J_\mu=v\partial_\mu a+\frac{x^{-1}}{x+x^{-1}}\sum_{i}\bar u_i\gamma_\mu
\gamma_5u_i+\frac{x}{x+x^{-1}}\sum_{i}\bar d_i\gamma_\mu\gamma_5d_i
+(\cdots)
\end{equation} 
where ($\cdots$) imply the contribution coming from the process
of $a,\pi^0,\eta$ diagonalization as given in Eq.~(20).

\subsection{the $a-\gamma-\gamma$ coupling}

This feeble coupling of axions can be probed by Sikivie's cavity 
experiments.\cite{sikivie} From the fundamental theory, one can
calculate the $a\gamma\gamma$ coupling, $\bar c_{a\gamma\gamma}$.
When going through the chiral symmetry breaking, it obtains an
additional contribution, and the total coupling $c_{a\gamma\gamma}$
is given by
\begin{equation}
c_{a\gamma\gamma}=\bar c_{a\gamma\gamma}-\frac{2}{3}\frac{4+Z}{1+Z}
=\bar c_{a\gamma\gamma}-1.92
\end{equation}
for $Z=0.6$, and
\begin{equation}
\bar c_{a\gamma\gamma}=\frac{E}{C}, \ \ E={\rm Tr}Q^2_{\rm em},\ \ 
\delta_{ab}C={\rm Tr}\lambda_a\lambda_bQ_{\rm PQ}.
\end{equation}
Note that for the KSVZ model, $C_3=-1/2, C_8=-3, E_3=-3e^2_Q$, and
$E_8=-8e^2_Q$. Thus, $\bar c_{a\gamma\gamma}=6e_Q^2$ and $(8/3)e^2_Q$
for color triplet and octet quarks, respectively. For the DFSZ model, $
C_3=N_g, E=(8/3)N_g$ where $N_g$ is the family number. The DFSZ model
is distinguished how electron obtains mass. If it gets mass through
$H_1$, $\tilde H_2$, and $H_3$ coupling, respectively, 
then they can be called
$(d^c,e)$--, $(u^c,e)$--, and non--unification models, respectively.
These couplings are listed in Table 1. 

\vskip 0.15cm
\centerline{Table 1. $c_{a\gamma\gamma}$ for several KSVZ and DFSZ
models.}
\begin{center}
\begin{tabular}{|cc|cc|}
\hline
\ \ \ \ KSVZ& & DFSZ &\\
$e_R$ & \ \ \ $c_{a\gamma\gamma}$ & $x$ (unif)  & 
\ \ \ $c_{a\gamma\gamma}$\\
\hline
$e_R=0$ & \ \ \ \ --1.92 & any ($d^c$) & \ \ \ \ 0.75\\
$e_3=-1/3$ & \ \ \ \ --1.25 & 1 ($u^c$) & \ \ \ \ --2.17\\
$e_3=2/3$ & \ \ \ \ 0.75 & 1.5 ($u^c$) & \ \ \ \ --2.56\\
$e_3=1$ & \ \ \ \ 4.08 & 60 ($u^c$) & \ \ \ \ --3.17\\
$e_8=1$ & \ \ \ \ 0.75 & 1 (non) & \ \ \ \ --0.25\\
$(m,m)$ &\ \ \ \  --0.25 & 1.5 (non) &\ \ \ \ --0.64\\
$(1,2)$&\ \ \ \ --0.59& 60 (non) & \ \ \ \ --1.25\\
\hline
\end{tabular}
\end{center}

\vskip 0.15cm
Under the assumption that these very light axions 
are the dark matter of our galaxy, we can predict the
detection rate in the Sikivie type detector. The model
predictions and the experimental bounds are compared 
in Fig. 1. In reality, there can be many heavy quarks which carry
nontrivial PQ charges. There are also two Higgs doublets in
supersymmetric models. For example, in superstring models, there
are more than 400 chiral fields, in which there appears in most
cases heavy quarks. Therefore, if the PQ symmetry is spontaneously
broken in superstring models, then it must contain both the KSVZ
and the DFSZ aspects. If a string model is given, one can calculate
$c_{a\gamma\gamma}$.

\begin{figure}
\epsfxsize=160mm
\centerline{\epsfbox{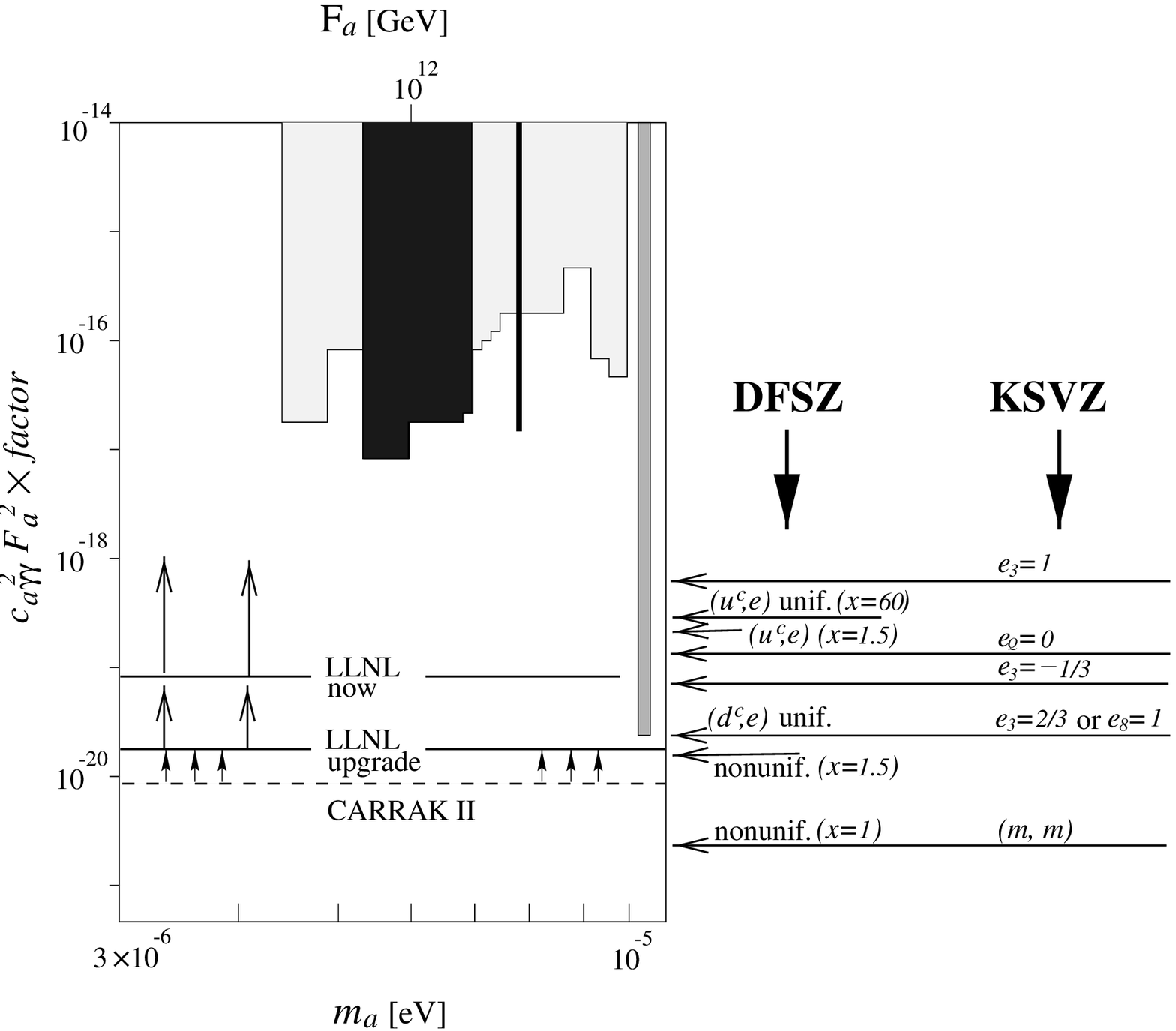}}
\vskip 0.5cm
\centerline{Fig. 1. Comparison of several $c_{a\gamma\gamma}$'s
and experimental bounds.} 
\end{figure}

\section{Astrophysical and Cosmological Bounds}

The astrophysical and cosmological bounds allow a window
of $F_a$,
\begin{equation}
10^{9-10}\ {\rm GeV}\le F_a\le 10^{12}\ {\rm GeV}.
\end{equation}
The astrophysical bounds come from the requirement that the
axions produced at the core of a star do not take out too
much energy compared to the takeout by neutrinos. If the
axion coupling is too small, the axions are not produced
copiously in the star and hence even though they do not have a chance
to be blocked by matter in the star, it will be allowed.
This condition translates to the lower bound on
the axion decay constant. The most stringent bound comes 
from SN1987A.\cite{seckel} The upper bound around $F_a\simeq
10^{12}$~GeV  comes from the cold axion energy density of the 
universe.\cite{pww} 

But it is known that the topological defects such as 
axionic strings~\cite{shellard} which is comparable to the
cold axion energy density and axionic domain walls~\cite{domain} 
which overcloses the universe. However, 
these will not cause a serious
problem in the inflationary universe paradigm. With supergravity
inflation, anyway one needs the reheating temperature after
inflation less than $10^9$~GeV~\cite{ekn} which is
supposed to be less than $F_a$.
Thus the most robust bound on cosmological bound of $F_a$
comes from the cold axion energy density which is~\cite{turn}  
\begin{equation}
\Omega_ah^2=0.13\times 10^{\pm 0.4}\Lambda_{200}^{-0.7}
f(\theta_1)m_{a5}^{-1.18}N_{\rm DW}^2	
\end{equation}
where $\Lambda_{200}$ is the QCD scale in units of 200~MeV,
$f(\theta_1)$ is the initial value if $\bar\theta$ at the
starting moment of the axion rolling around $T\sim 1$~GeV,
and $m_{a5}=m_a/10^{-5}$~eV. Note that superstring models have
$N_{\rm DW}=1$.\cite{wit} In general, the $F_a$ bound is
customarily given as in Eq.~(25).\footnote{For a high reheating
temperature, Battye and Shellard \cite{bs} 
give $F_a<4\times 10^{10}$~GeV, but
the Harari-Sikivie\cite{shellard} estimate of the axion string energy is
smaller than the Battye-Shellard estimate by a factor 10--100.}
 If $\Omega_a\simeq 0.3$,
then we have a bound $F_a\simeq 5\times 10^{11}$~GeV and
$m_a\simeq 1.2\times 10^{-5}$~eV. Assuming that the cold dark 
matter in our galaxy is mostly the cold axion, the axion search
experiment is going on, which was reviewed above.

\section{Superstring Axion}

The standard introduction of axion through spontaneous 
symmetry breaking of the Peccei-Quinn global symmetry is
ad hoc. As we have seen, there are many ways to introduce
axions. Among these, there exists a very interesting method.
It is from string theory. Here, the very light axion must
be present always.\cite{witten} Furthermore, the string
theory--the anticipated theory of everything--must solve the
strong CP problem if it is a physical theory. Therefore, it
is worthwhile to see what is the problem in superstring axions.  

The 10D string theory contains the bosonic degrees,$G_{MN}, B_{MN}$
and $\phi$ where $\{M,N\}=0,1,2,\cdots,9$. $G_{MN}$ contains
the graviton, $\phi$ is called the dilaton, and the 
antisymmetric tensor $B_{MN}$ contains the axion.
When $M$ and $N$ are restricted to 4D indices $\mu,\nu,$ etc.,
it has one physical degree.  The 4D field strength of $B$ is
denoted as $H$, the three index antisymmetric tensor. The
axion is the dual of $H$,
\begin{equation}
\partial^\sigma a\propto \epsilon^{\mu\nu\rho\sigma}H_{\mu\nu\rho},
\ \ H_{\mu\nu\rho}\propto \epsilon_{\mu\nu\rho\sigma}
\partial^\sigma a.
\end{equation}
Why can we call this axion? It is because it has a coupling
given in the form of Eq.~(2). Let us see how it arises.

The field strength $H$ is not just the curl of $B$, but is
made gauge invariant by adding Chern-Simmons terms,
\begin{equation}
H=dB-\omega^0_{3Y}+\omega^0_{3L}
\end{equation}
where the Yang-Mills and Lorentz Chern-Simmons forms satisfy
$d\omega^0_{3Y}={\rm tr}F^2$ and $d\omega^0_{3L}={\rm tr}R^2$.
Therefore,
\begin{equation}
dH=-{\rm tr}F^2+{\rm tr}R^2.
\end{equation}
The reason, $B$ transforms nontrivially under gauge transformation
to cancel the Yang-Mills and Lorentz anomalies,\cite{green} is
the source of Eq.~(29). Also, the Green-Schwarz term contains
\begin{equation}
S_{\rm GS}\propto \int (B{\rm tr}F^4+\cdots).
\end{equation} 
Eq.~(29), through the duality relation (27), gives the following
equation of motion for $a$,
\begin{equation}
\partial^2 a=-\frac{1}{M_c}({\rm Tr}F_{\mu\nu}
\tilde F^{\mu\nu}-{\rm Tr}R_{\mu\nu}\tilde R^{\mu\nu})
\end{equation}
which can be obtained from an effective Lagrangian
\begin{equation}
{\cal L}=\frac{1}{2}(\partial_\mu a)^2 -
\frac{a}{M_c}({\rm Tr}F\tilde F-{\rm Tr}R\tilde R)
\end{equation}
where $M_c$ is the compactification scale.
This axion is called the {\it model-independent axion} (MIa). Any
string models have this one.

Eq.~(30) also has a coupling which allows to interpret another 
pseudoscalar $a^\prime$ an axion. It depends on the compactification 
scheme.\cite{choi} This model-dependent axion receives contributions
to the superpotential from the world-sheet instanton effects, obtaining
a large mass in general.\cite{wen} Therefore, it is not useful
for the solution of the strong CP problem. Recently, it has been
pointed out that in the M-theory this model-dependent axion
can have a room to play for the strong CP solution.\cite{choi1}

In sum, there are axions in string models. The model-independent
axion is the promising one toward a solution of the strong CP
problem. But it has two problems to be solved:
\vskip 0.15cm
\indent (A)  Decay constant problem--The MIa has the decay constant
around $F_a\simeq 10^{16}$~GeV~\cite{ck} which is too large for 
the axion cold dark matter scenario.\\
\indent (B) Hidden sector problem--Popular supergravity models require
a confining hidden sector, say $SU(N)_h$, at $\sim 10^{13}$~GeV scale.
If so, MIa obtains a mass also from the hidden sector contribution.
Because the hidden sector potential is so steep compared to the QCD
potential that it will settle $\theta_h\simeq 0$, but leaves $\theta_{\rm
QCD}\ne 0$. Therefore, one axion cannot solve the strong CP problem. 
Two axions are needed with the hidden sector.
\vskip 0.15cm

Problem (A) seems to be solved in anomalous $U(1)$ models.\cite{anom} 
In this case, the MIa becomes the longitudinal degree of the
anomalous gauge boson. The gauge boson and MIa is removed at
this gauge boson mass scale. At low energy, there remains a
global PQ symmetry which can be broken at $\sim 10^{12}$~GeV.
However, if the low energy theory maintains supersymmetry,
it is very difficult to realize this scenario. The reason is that
the supersymmetry condition with matter fields 
always breaks the global PQ symmetry
at the gauge boson mass scale, and a remaining $U(1)$ symmetry
(if any) turns out to be a gauge symmetry.

To solve (B), we need another axion or massless hidden sector
fermion. One obvious choice is massless hidden sector gluino
(h-gluino). The h-gluino will obtain mass eventually; but 
this case is different from the massless up quark case since
h-gluino mass is generated by the h-gluino condensation.
Massless h-gluino suggests a possible global symmetry; $U(1)_R$
symmetry. But there is no global symmetry in string models
except that corresponding to the MIa.\cite{banks} Therefore,
the best we can do is to consider a discrete subgroup of $U(1)_R$
when we consider the massless h-gluino. Under certain assumptions,
it is easy to calculate the axion dependence of the potential. 
For $Z_N$ subgroup of $U(1)_R$, we have $V\sim 10^{13\sim -25}$~GeV$^4$
for $N=2$, $10^{-29\sim -8}$~GeV$^4$ for $N=3$, $10^{-8\sim
10}$~GeV$^4$ for $N=4$, $10^{-50\sim -26}$~GeV$^4$, etc. Except
$Z_2$ and $Z_4$, any discrete subgroup of $U(1)_R$ is phenomenologically
acceptable.\cite{gkn} The schematic behavior of the contributions
to the potential is shown below.

\font\eightrm=cmr8

\noindent
In general, we expect the following contributions to $V$,
%
%
\begin{figure}[htb]
$$\beginpicture
\setcoordinatesystem units <72pt,18pt> point at 0 0
\setplotarea x from -0.700 to 2.100, y from 0.000 to 1.500
\setplotsymbol ({\normalsize.})
\setsolid
\setquadratic
\put {QCD} [c] at -0.200 2.600 
\put {SU(N)$_{\rm h}$} [c] at 1.600 2.600 
\plot
-0.700  0.2   -0.650  0.291 -0.600  0.362 -0.550  0.398 -0.500  0.390 
-0.450  0.341 -0.400  0.262 -0.350  0.169 -0.300  0.082 -0.250  0.022
-0.200  0.000 -0.150  0.022 -0.100  0.082 -0.050  0.169  0.000  0.262
 0.050  0.341  0.100  0.390  0.150  0.398  0.200  0.362  0.250  0.291
 0.300  0.2  
/
\put {$\bullet$} <0pt,2pt> at  0.050 0.341 
\plot
 1.1    1.000  1.15   1.454  1.200  1.809  1.250  1.988  1.300  1.951 
 1.350  1.707  1.400  1.309  1.450  0.844  1.500  0.412  1.550  0.109
 1.600  0.000  1.650  0.109  1.700  0.412  1.750  0.844  1.800  1.309
 1.850  1.707  1.900  1.951  1.950  1.988  2.000  1.809  2.050  1.545
 2.1    1.000    
/
\put {$\bullet$} <0pt,2pt> at 1.600 0.000
\endpicture$$
\end{figure}

\noindent
But with a massless h-sector gluino, the contributions will look like 
%
%
\begin{figure}[htb]
$$\beginpicture
\setcoordinatesystem units <72pt,18pt> point at 0 0
\setplotarea x from -0.700 to 2.100, y from 0.000 to 1.000
\setplotsymbol ({\normalsize.})
\setsolid
\setquadratic
\put {QCD} [c] at -0.200 2.600 
\put {SU(N)$_{\rm h}$} [c] at 1.600 2.600
\plot
-0.700  1.000 -0.650  1.454 -0.600  1.809 -0.550  1.988 -0.500  1.951
-0.450  1.707 -0.400  1.309 -0.350  0.844 -0.300  0.412 -0.250  0.109
-0.200  0.000 -0.150  0.109 -0.100  0.412 -0.050  0.844  0.000  1.309
 0.050  1.707  0.100  1.951  0.150  1.988  0.200  1.809  0.250  1.454
 0.300  1.000  
/
\put {$\bullet$} <0pt,2pt> at -0.200 0.000
\setlinear 
\plot
1.100 0.900 2.100 0.900
/
\put {$\bullet$} <0pt,2pt> at 1.600 0.900
\endpicture$$
\end{figure}

\noindent
With sufficiently suppressed h-sector contribution ($Z_3, Z_5$, etc.), 
$\theta_{\rm QCD}$ is settle at 0 but $\theta_h\ne 0$,
%
%
\begin{figure}[htb]
$$\beginpicture
\setcoordinatesystem units <72pt,18pt> point at 0 0
\setplotarea x from -0.700 to 2.100, y from 0.000 to 1.000
\setplotsymbol ({\normalsize.})
\setsolid
\setquadratic
\put {QCD} [c] at -0.200 2.600 
\put {SU(N)$_{\rm h}$} [c] at 1.600 2.600 
\plot
-0.700  1.000 -0.650  1.454 -0.600  1.809 -0.550  1.988 -0.500  1.951
-0.450  1.707 -0.400  1.309 -0.350  0.844 -0.300  0.412 -0.250  0.109
-0.200  0.000 -0.150  0.109 -0.100  0.412 -0.050  0.844  0.000  1.309
 0.050  1.707  0.100  1.951  0.150  1.988  0.200  1.809  0.250  1.454
 0.300  1.000  
/
\put {$\bullet$} <0pt,3pt> at -0.200 0.000
\plot
 1.1    0.2    1.15   0.291  1.200  0.362  1.250  0.398  1.300  0.390
 1.350  0.341  1.400  0.262  1.450  0.169  1.500  0.082  1.550  0.022
 1.600  0.000  1.650  0.022  1.700  0.082  1.750  0.169  1.800  0.262
 1.850  0.341  1.900  0.390  1.950  0.398  2.000  0.362  2.050  0.291
 2.1    0.2    
/
\put {$\bullet$} <0pt,2pt> at 1.850 0.341
\endpicture$$
\end{figure}

\noindent Therefore, the QCD axion physics is the usual one, satisfying
the condition for the cold dark matter.
[{\it Note added after the talk:} In fact, if this scenario 
of making the hidden sector potential
extremely shallow, one can obtain a reasonable value for a
nonvanishing cosmological constant.\cite{lambda} The magnitude of
the cosmological constant is roughly the height shown in the
last cartoon.]

\section{Conclusion}

I reviewed briefly that the strong CP solution is guaranteed with
a very light axion, its possible imbedding in superstring models,
cosmological implication as a cold dark matter candidate, and
possible detection of this very light axion in the
cavity detectors assuming that the missing mass is mostly 
these axions. If discovered, its implication will be dramatic,
since it will prove experimentally the whole idea of instanton,
the collective axion oscillation, and maybe the superstring
idea. On the other hand, if not discovered, the axion misalignment
angle may be very small so that there are not so much axions
as needed by the missing mass, or the axion decay constant may
be very large.\cite{choi1,pi} Also, the strong CP problem must
be solved outside the axion mechanism, which is however considered
to be not so attractive.\cite{rev}

\section*{Acknowledgments}
This work is supported in part by KOSEF,  MOE through
BSRI 98-2468, and Korea Research Foundation.

\end{document}